\documentclass[conference]{IEEEtran}

\usepackage{amsmath}
\usepackage{amssymb}
\usepackage{cite}
\usepackage{psfrag}
\usepackage{graphicx}
\usepackage{dsfont}
\usepackage{epsfig}

\IEEEoverridecommandlockouts
\newcommand{\mb}[1]{\mathbf{#1}}

\begin{document}
%
\title{Annotated Raptor Codes}

\author{\IEEEauthorblockN{Kaveh Mahdaviani, Masoud Ardakani, Chintha Tellambura}
\IEEEauthorblockA{Department of Electrical and Computer Engineering\\
University of Alberta, Edmonton, AB, Canada. T6G 2V4\\
Email: {mahdaviani,ardakani,tellambura}@ece.ualberta.ca}}

%


\maketitle

\begin{abstract}
In this paper, an extension of raptor codes is introduced which keeps all the desirable properties of raptor codes, including the linear complexity of encoding and decoding per information bit, unchanged. The new design, however, improves the performance in terms of the reception rate. Our simulations show a 10\% reduction in the needed overhead at the benchmark block length of 64,520 bits and with the same complexity per information bit.
\end{abstract}

\section{Introduction}
Fountain codes such as LT code \cite{LT} and raptor codes \cite{Raptor} were originally designed to achieve the capacity on any binary erasure channel (BEC) with no channel information and at very low complexity. The decoding complexity of raptor codes under edge deletion (ED) decoding \cite{LT, Raptor} is linear with the block length. Therefore, these codes are the natural choice for data transmission over channels with unknown or very fast changing properties. Raptor codes preserve many of their interesting properties over other channels such as the binary symmetric channel, additive white Gaussian noise and fading channels \cite{Raptor_Design_BinSym,Raptor_on_BSC,Raptor_on_AWGN,Raptor_on_Fading}. These codes have been already adapted as the forward error correction code for multimedia broadcast/multicast services (MBMS) by the 3rd Generation Partnership Project (3GPP) \cite{3GPP}.

In practice, it is well known that ED needs a small overhead for successful decoding. Specifically, to decode $k$ information bits, $k (1 +\varepsilon)$ received bits are needed at the decoder, where $\varepsilon$ is referred to as the overhead. More specifically, even with the highly optimized designs a nonzero overhead is needed if using the low complexity ED decoding. In order to avoid this overhead or reduce it to a negligible amount, a more complex decoding algorithm is introduced for raptor codes called the inactivation decoding \cite{Inact_Decode}, but this algorithm is practical only for small block lengths due to its nonlinear complexity. In this paper, we introduce a variation of raptor codes called the annotated raptor codes which reduce the overhead of conventional raptor codes while keeping the encoding and decoding complexity linear. Although design of these codes is out of the scope of this short paper, numerical examples are provided to demonstrate their lower overhead even without a fine optimization.

After a quick review of conventional raptor codes and introducing our notations in Section II, in Section III we provide our main idea for reducing the overhead while keeping the complexity unchanged. In Section IV, we describe the encoding and decoding of the proposed annotated raptor codes which is continued by some general comments on the design of code parameters in Section V. Finally in Section VI, a numerical example on a benchmark block length is presented. The paper is concluded in Section VII.




\section{Background and Notations}

In this section, we briefly review the encoding and decoding of conventional raptor codes. Unlike what is most common in the literature of rateless codes, we use the matrix form rather than the graph representation. The matrix form is more suitable for explaining annotated raptor codes later. In this section, we also introduce the notations and definitions that will be used later.

\subsection{Encoding}
The encoding starts with a fixed rate outer code of rate $R$ and a parity check matrix $\mb{H}_{(n-k)\times n}$ which encodes an information block of $k$ input bits into a block of $n=\frac{k}{R}$ encoded bits $b_{1},\ldots,b_{n}$, called the intermediate bits. To produce an output bit, first, the encoder randomly samples an integer $m \in \{1,\ldots,D\},~D\leq n$ from a probability distribution. This distribution is characterized by a generating polynomial \[\Omega(x)=\sum_{i=1}^{D}{\Omega_{i}x^{i}}\] where $\Omega_{i}$ is the probability that $m=i$. The encoder then uniformly chooses a set of $m$ intermediate bits and produces an output bit by XORing them. Output bits are produced and transmitted until enough bits are received by the decoder to recover the information bits successfully.

Each output bit can be viewed as a parity check equation on a subset of intermediate bits, where the parity value is transmitted on the channel. 
The outer code can also be viewed as a set of parity check equations on intermediate bits. Unlike before, these parity values are always zero, thus they need not be transmitted on the channel. The decoder can use all the outer code equations and any received output bit equation to form an equation system from which all the intermediate bits are recovered. The information bits are then obtained through a linear mapping from intermediate bits according to the outer code.

\subsection{Edge Deletion Decoding}

The decoder starts with a linear equation system consisting of the parity check equations of the outer code
\begin{align}
\mb{HX}=\mb{O}_{(n-k)\times 1} \nonumber
\end{align}
where $\mb{O}_{\ell \times k}$ represents an $\ell \times k$ all zero matrix. At this point the set of recovered intermediate bits is still empty. Assuming the BEC with erasure rate $\delta$, with probability $1-\delta$ an output bit is received. Receiving each output bit $b_{i}$ enables the decoder to use $b_i$'s corresponding parity check equation.

Upon receiving an equation, the decoder will substitutes any recovered intermediate bits, and then adds the reduced equation to its equation system. Whenever a reduced equation is of weight one, the equation is put in a set called the ripple. For any equation in the ripple, the value of the intermediate bit is immediately known and can be substituted in every other equation. This procedure is called the elimination process. It is easy to check that the order of using ripple elements have no effect on the performance of the decoder. Note that during the elimination process, the weight of some of the rows of the coefficient matrix is reduced which could in turn result in achieving new equations of weight one, and refilling the ripple. If the ripple gets empty before all the intermediate bits are recovered, receiver will listen to the channel to receive more equations to refill the ripple.

After receiving enough bits for a successful ED decoding, we have the following linear equation system.
\begin{align} \label{initial-ED}
\left[\begin{array}{c}
\mb{H}_{(n-k)\times n}\\
\mb{C}_{(1+\varepsilon)k\times n}
\end{array}\right] \mb{X} =
\left[\begin{array}{c}
\mb{O}_{(n-k)\times 1} \\
\mb{R}_{(1+\varepsilon) k \times 1}
\end{array}\right]
\end{align}
where, $\mb{C}$ is the coefficient matrix of the parity check equations corresponding to the received bits, $\varepsilon$ is the overhead, $\mb{R}$ is the vector containing the value of the received bits and $\mb{X}$ is the set of unknown intermediate variables. After successfully finishing ED decoding, upon reordering the rows, we obtain the following matrix equation.
\begin{align} \label{finished-ED}
\left[\begin{array}{c}
\mb{I}_{n} \\
\mb{O}_{\varepsilon k \times n}
\end{array}\right]\mb{X}=
\left[\begin{array}{c}
\mb{B}_{n\times 1} \\
\mb{O}_{\varepsilon k \times 1}
\end{array}\right].
\end{align}
Here, $\mb{B}=[b_{1},\ldots,b_{n}]^T$ is a vector, containing the recovered values of the intermediate bits. Note that the reordering is just required for simplifying the representation. In the real implementation, this reordering is not needed.

\section{Main Idea}
According to \cite{Raptor}, even using highly optimized raptor codes with very large block lengths, the overhead is nonzero. This means that some of the received bits will be useless. These received bits are represented as all zero rows at the end of ED decoding (see Eq. (\ref{finished-ED})). In other words, these rows contain no new information about the intermediate bits given the rest of equations, thus we call them the ``overhead rows''. Although, it is not possible to avoid the overhead rows, interestingly, we will see that it is still possible to embed new information bits in them.

To embed new information bits in overhead rows, we first add an auxiliary set of variables $a_{1},\ldots,a_{n_a}$ to the binary equation system and extend the columns of the coefficients matrix. We refer to these auxiliary columns of the coefficient matrix and their corresponding set of variables as ``A-columns'' and ``A-variables'' respectively. Clearly, the A-columns are not all zeros. Thus, some of the output bits are now XORed with bits from the A-variables. We refer to this operation as ``annotation''. The details of this operation is presented in Section \ref{Annotated-Enc}. As we will explain later, the A-variables themselves must be protected by a low-rate outer code. Let us denote the $(n_a-k_a)\times n_a$ parity check matrix of this outer code by $\mb{H}^{(a)}$ and the encoded block by $\mb{A}=[a_{1},\dots,a_{n_a}]^T.$

As a result, in the decoding process the initial matrix form represented in Eq. (\ref{initial-ED}) changes to
\begin{align}
&\left[\begin{array}{c|c}
\mb{H}_{(n-k)\times n}&\mb{O}_{(n-k)\times n_a}\\
\mb{O}_{(n_a-k_a)\times n}&\mb{H}^{(a)}_{(n_a-k_a)\times n_a}\\
\mb{C}_{(1+\varepsilon')(k+k_a)\times n}&\mb{C}^{(a)}_{(1+\varepsilon')(k+k_a)\times n_a}
\end{array}\right]
\left[\begin{array}{c}
\hspace{2mm}\\ \mb{X}_{n\times 1}\\ \hspace{2mm}\\
\cline{1-1} \mb{X}^{(a)}_{n_a\times 1}
\end{array}\right]\nonumber
 =\\
&\left[\begin{array}{c}
\mb{O}_{(n+n_a-(k+k_a))\times 1} \\
\mb{R}_{(1+\varepsilon') (k+k_a) \times 1}
\end{array}\right]
\end{align}
In the above equations $[\mb{C}|\mb{C}^{(a)}]$ is the coefficient matrix of the parity check equations corresponding to the received bits where, $\mb{C}$ part represents the coefficients of the intermediate bits and $\mb{C}^{(a)}$ represents the coefficients of the annotation bits. Notice that $n_a$ extra intermediate bits, which carry $k_a$ new information bits, are now added to the system. Thus, $\varepsilon'$ represents the new overhead. Finally upon reordering of rows the final form after successful ED decoding is
\begin{align}
\left[\begin{array}{c}
\begin{array}{c|c}
\mb{I}_{n}&\mb{O}_{n\times n_a}\\
\mb{O}_{n_a\times n}&\mb{I}_{n_a}\\
\end{array}\\
\cline{1-1}
\mb{O}_{\varepsilon'(k+k_a)\times (n+n_a)}
\end{array}\right]
\left[\begin{array}{c}
\hspace{2mm}\\ \mb{X}_{n\times 1}\\ \hspace{2mm}\\
\cline{1-1} \mb{X}^{(a)}_{n_a\times 1}
\end{array}\right]\nonumber
=\left[\begin{array}{c}
\mb{B}_{n \times 1} \\
\mb{A}_{n_a \times 1} \\
\mb{O}_{\varepsilon'(k+k_a) \times 1}
\end{array}\right]
\end{align}


The details of ED decoding for an annotated equation system is provided in Section \ref{Annotated-Dec}. Here, to make the main idea more clear, we present a toy example. Assume that we have a block of three bits $x_{1},x_{2},x_{3}$, and we produce output symbols of degrees 1 to 3 with equal probabilities. Now, if for example the receiver receives $r_{1}=x_{1}\oplus x_{2},~r_{2}=x_{1}\oplus x_{2}\oplus x_{3},~r_{3}=x_{2}\oplus x_{3},$ and $r_{4}=x_{1}$, then the ED decoding of intermediate bits will not perform any elimination process before receiving $r_{4}$. When $r_{4}$ is received, it goes to the ripple and ED decoding starts recovering the values of intermediate bits. The received equation system before performing elimination is
\begin{align}
\left[\begin{array}{ccc}
1&1&0\\
1&1&1\\
0&1&1\\
1&0&0\\
\end{array}\right]
\left[\begin{array}{c}x_{1}\\x_{2}\\x_{3}\end{array}\right]&= \nonumber
\left[\begin{array}{c}r_{1}\\r_{2}\\r_{3}\\r_{4}\end{array}\right]
\end{align}

It is easy to check that ED decoding will recover all the intermediate bits with this equation system and after ED decoding we have
\begin{align}
\left[\begin{array}{ccc}
0&1&0\\
0&0&1\\
0&0&0\\
1&0&0\\
\end{array}\right]
\left[\begin{array}{c}x_{1}\\x_{2}\\x_{3}\end{array}\right]&= \nonumber
\left[\begin{array}{c}r_{1}\oplus r_{4}\\r_{1}\oplus r_{2}\\r_{2}\oplus r_{3}\oplus r_{4}\\r_{4}\end{array}\right]
\end{align}

In the above, obviously, the third row is an overhead row and contains no new information about the intermediate bits given all the other rows. But if we annotate some of the transmitted bits (say $r_2$, $r_3$ and $r_4$) with a single A-variable $a$, then the representation of equation system after receiving $r_{4}$ is
\begin{align}
\left[\begin{array}{ccc|c}
1&1&0&1\\
1&1&1&1\\
0&1&1&1\\
1&0&0&1\\
\end{array}\right]
\left[\begin{array}{c}x_{1}\\x_{2}\\x_{3}\\ \cline{1-1} a \end{array}\right]&= \nonumber
\left[\begin{array}{c}r_{1}\\r_{2}\\r_{3}\\r_{4}\end{array}\right]
\end{align}

The ED decoding can start the elimination and recovery procedure at this point, if we perform the decoding only based on the intermediate bits and in terms of the annotated variable $a$. As a result, when the ED decoding of intermediate bits finishes, the resulting equation system has the following form
\begin{align}
\left[\begin{array}{ccc|c}
0&1&0&0\\
0&0&1&0\\
0&0&0&1\\
1&0&0&1\\
\end{array}\right]
\left[\begin{array}{c}x_{1}\\x_{2}\\x_{3}\\ \cline{1-1} a \end{array}\right]&= \nonumber
\left[\begin{array}{c}r_{1}\oplus r_{4}\\r_{1}\oplus r_{2}\\r_{2}\oplus r_{3}\oplus r_{4}\\r_{4}\end{array}\right]
\nonumber
\end{align}
Notice that still the third equation does not play any role in the recovery of the intermediate bits, but this row can be used to recover the value of the A-variable $a$ as $a=r_2 \oplus r_3\oplus r_4$. The A-variable can in turn be used to recover any intermediate bit which was computed in terms of the A-variable (in this case $x_1$ in the fourth row). This example shows that with the same number of received bits, it is possible to recover more intermediate bits using annotation. Of course this was a highly fabricated example, in which the overhead was reduced to zero. Clearly, we do not expect zero overhead in a practical setup. However, as will be seen, the annotation idea retrieves a portion of the overhead at no extra cost. In fact, in order to keep the decoding complexity unchanged per information bit, we will see that the decoding procedure used in this toy example is not desirable. In Section \ref{Annotated-Dec} we propose a revised version of ED decoding for annotated raptor codes.

\section{Annotated Raptor Codes}

Ideally we prefer to perform the annotation such that it will not affect any of the desirable properties of the original raptor codes. More specifically, we do not want to increase the complexity per bit (neither at the encoder nor at the decoder). Achieving this goal, however, requires careful annotation and decoding. To see why the trivial approach (similar to the one in the toy example above) may fail, note that when the ED decoder uses annotated rows as pivots in row operations, extra complexity is resulted from the 1's in the corresponding rows of the A-columns. Thus a high-density of 1's in the A-columns is against the goal of a low-complexity design. Unfortunately, even starting with sparse A-columns, the density of 1's in the A-columns gradually increases as ED decoding progresses. Our numerical simulation shows that the complexity will grow super linear with an approximate exponent of $1.3$. In the following, we briefly outline an annotation method that achieves linear complexity.

Let us assume that we could know beforehand which transmissions end up as overhead rows. If this knowledge existed, we could annotate only these transmissions. Although such a knowledge cannot exist in a real setup, we can annotate a small portion of rows and pretend that they will end up being the overhead rows. Thus, the decoding will start from the non-annotated rows. Our interesting observation is that if annotated rows are selected carefully, ED decoding of non-annotated rows will recover a large portion of intermediate bits. In other words, assume in a conventional raptor code, we carefully select and mark a $\sigma_{0}$ portion of the transmissions for annotation. Then, in the receiver, we first exclude the marked received bits and perform ED decoding on the unmarked received equations and the parity check equations of the outer code. When the total number of received bits is close to the number of input bits, we observe that the decoder recovers a $(1-\delta_{0})$ portion of the intermediate bits. Typically for $\sigma_0=0.05,$ we have $\delta_0=0.3.$

After recovery of $(1-\delta_{0})$ portion of the intermediate bits, it is easy to see that with probability $(1-\delta_{0})^{i}$, an annotated equation which originally contains $i$  intermediate bits, is reduced to an equation based only on the A-variables. We call these reduced equations the ``A-equations''. From these A-equations a fixed portion of A-variables will be recoverable. Now, if the rate of the outer code of the A-variables is selected properly, it will be possible to decode all the A-variables. Consequently, it will be possible to de-annotate all the annotated rows in linear complexity. In fact to keep the complexity of this de-annotation at its absolute minimum, in this work, each annotated row has a single A-variable in it. This also keeps the encoding complexity linear.

Finally, after de-annotation, the rest of the intermediate bits will be recovered using ED decoding. In terms of ED decoding of the intermediate bits, the only difference between an annotated raptor code and a conventional one is that here we have changed the order of using the received equations. We use some of the equations at first and postpone using the others (the annotated ones) for a while. Between these two phases, we recover some information bits that are embedded in the annotation. 

In the next section we will go through more details of the encoding and decoding algorithms for the annotated raptor codes.

\subsection{Encoding} \label{Annotated-Enc}

The encoding process in annotated raptor codes has two separate steps. In the first step, two information blocks of length $k$ and $k_{a}$ are coded into two encoded blocks (i.e., the intermediate variables and the A-variables), using fixed rate outer codes with parity check matrices $\mb{H}_{(n-k)\times n}$, and $\mb{H}^{(a)}_{(n_{a}-k_{a})\times n_{a}}$.

The second step, which contains two phases, will generate an output bit. First an integer $m\in \{1,\ldots,D\},~D\leq n$ will be sampled based on a probability distribution represented by its generating polynomial \[\Theta(x)=\sum_{i=1}^{D}{(\Phi_{i}+\Psi_{i})x^{i}}.\] Here $m = i$ happens with probability $(\Phi_{i}+\Psi_{i})$. Consequently, based on the selected value of $m$, encoder samples another random variable $b\sim \mathcal{B}\left(\frac{\Psi_{m}}{\Phi_{m}+\Psi_{m}}\right)$, where $\mathcal{B}\left(p\right)$ represents the Bernoulli distribution with probability of success equal to $p$. The encoder then chooses $m$ intermediate bits uniformly at random. If the Bernoulli outcome is success, a single A-variable bit is also selected uniformly at random. Finally, the XOR of all the selected bits forms an output bit for transmission. Output bits are generated and transmitted iteratively, until successful transmission of the whole data block.

\subsection{Decoding} \label{Annotated-Dec}

The decoding procedure has already been described earlier in this section. Here we summarize the procedure. Two separate edge deletion decoders are used. The first one decodes the intermediate bits, using the non-annotated equations and the rows of matrix $\mb{H}$. The second one decodes the A-variables using any row whose first $n$ elements are all zeros including the rows of matrix $\mb{H}^{(a)}$. Obviously, as the decoders recover some of the intermediate bits and A-variables they remove them from all the equations and hence each decoder may provide the other with some new equations to be used in the rest of the decoding process. When both the decoders run out of ripple, receiver listens to the channel to receive new equations and refill at least one of the ripples again.

\section{Some Comments on Design}
Assume that the decoder has already received $n=(1+\varepsilon)k$ bits. Moreover, assume that through numerical search we have obtained the probability distribution $\Theta(x)$ for which, excluding the annotated received bits, ED decoding is able to recover a $\delta_{0}$ portion of the intermediate bits. Based on the previous discussions, the probability that a randomly selected row be reduced to an A-equation is
\begin{align}
P^{*}=\sum_{i=1}^{D}{\Psi_{i}(\delta_{0})^{i}}.\nonumber
\end{align}
Therefore, the average number of A-equations released by ED decoding of intermediate bits excluding annotated equations, is $(1+\varepsilon)kP^{*}$. According to the single-bit annotation strategy taken in this paper, the probability that a randomly selected A-variable is not covered in the released A-equations is
\begin{align}
(1-\frac{1}{n_a})^{((1+\varepsilon)kP^{*})}\simeq e^{(\frac{-(1+\varepsilon)kP^{*}}{n_a})}.\nonumber
\end{align}
Hence, the average number of A-variables which are now recovered is approximately
\begin{align} \label{A-recovery-rate}
m_a=n_a\left(1-e^{\frac{-(1+\varepsilon)kP^{*}}{n_a}}\right).
\end{align}

It is seen from (\ref{A-recovery-rate}) that $m_a$ is an increasing function of $n_a$ and that $m_a < (1+\varepsilon)kP^{*}$. Therefore, the new overhead $\varepsilon'$ can be found as
\begin{align}\label{varepsilon-star}
\varepsilon'=\frac{\varepsilon k-m_a}{k+m_a}.
\end{align}
It is easily seen that $\varepsilon'<\varepsilon$ as long as $m_a>0$ (i.e., $n_a>0$). Moreover, $\varepsilon'$ is a strictly decreasing function of $n_a$. It means that as the number of A-variables increases, more information bits can be transmitted using annotation, and thus, a larger portion of the overhead can be retrieved. As a result the lower the rate of the outer code for A-variables, the smaller the overhead will be. The improvement in the overhead, however, is bounded because $m_a$ saturates as a function of $n_a$ (see Eq. (\ref{A-recovery-rate})).

\begin{figure}
\centering
\resizebox{3.4in}{!}{\includegraphics[scale=0.5]{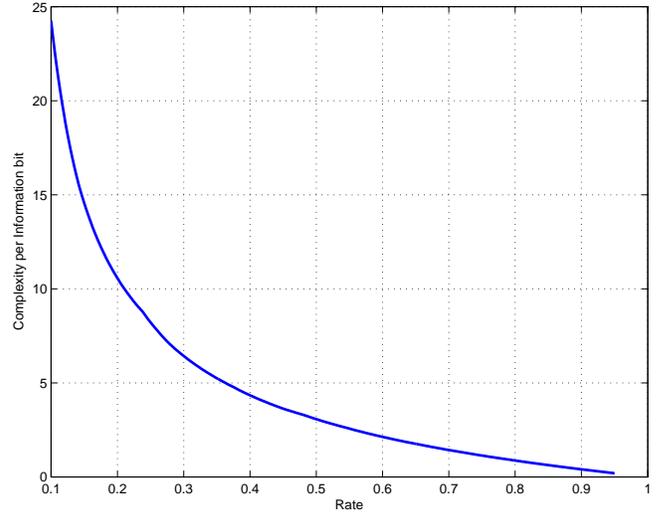}}\\
\caption{Complexity per information bit vs. rate for capacity approaching sequences of LDPC codes designed for the BEC.}
\label{fig:Complexity}
\end{figure}

A very low rate outer code, however, introduces a significant source of complexity. Although there exist very good low rate codes with linear complexity such as LDPC codes designed for erasure channels \cite{LDPC_linear_time,LDPC_Oswald,LDPC_Banihashemi}, when the rate of these codes tend to zero, the coefficient of the linear complexity tends to infinity. Figure \ref{fig:Complexity} depicts complexity per information bit, measured as the number of XORs needed for encoding/decoding of LDPC codes designed in \cite{LDPC_linear_time}. This figure is based on codes that achieve 95\% of the channel capacity.

To keep the complexity of annotated raptor codes equal to that of conventional raptor codes, we must use an outer code for the A-variables whose complexity per information bit is the same as conventional raptor codes. The complexity per information bit of a conventional raptor code is equal to the average weight of its output bits which is typically at least eight (considering the complexity of the high-rate outer code). Thus, Fig. \ref{fig:Complexity} suggests that the A-variables must be encoded using an outer code of rate around 0.25. Obviously, lower rate codes can be used to retrieve a higher portion of the overhead, but at the cost of a higher complexity per information bit. This extra complexity, however, is quite small since it affects only the parity check equations of the A-variables, which represent a small fraction of all equations (typically less than 4\%). Nonetheless, for any fixed rate outer code, the complexity remains linear.

Now assume we have selected an outer code of rate $R_a$ for A-variables which guarantees successful decoding of A-variables for erasure rates less than $1-R_a$ with high probability. According to the above discussions, we can now select the number of information bits $k_a$ to be encoded to $n_a$ A-variables as $k_a = n_a R_a$, where $n_a$ must satisfy
\begin{align}
R_a < 1-e^{(\frac{-(1+\varepsilon)kP^{*}}{n_a})}.\nonumber
\end{align}
Thus we have
\begin{align}\label{opt-k_a}
k_a < R_a \frac{-(1+\varepsilon)kP^{*}}{\ln(1-R_a)}.
\end{align}
This equation can be used to choose the number of information bits to be encoded by the rate $R_a$ outer code and be used as A-variables for annotations.

\section{Example Code and Numerical Results}
This section provides a numerical example of an annotated raptor code. As the optimization of the code is out of the scope of this paper, our example here does not represent an optimal design. Indeed, in order to better justify the benefits of annotated raptor codes, we focus on the impact of annotation on an existing probability distribution optimized for conventional raptor code. Clearly, we expect even better results through optimizing a probability distribution for annotated raptor codes.

Our focus in this example is on the highly optimized probability distribution $\Omega(x)$ presented in \cite{Raptor} for a raptor code with an information block of $k=$64,520 bits and an outer code of rate $R=0.9845$ to produce a block of $n=$65,536 intermediate bits. As mentioned before, we use a single bit annotation for the output bits that are selected to be annotated. This represents the simplest form of annotation. One may consider a degree distribution for the A-variables and optimize it for improved performance. Such optimizations, however, are out of the capacity of this paper.

Based on a set of numerical experiments we selected the probability distribution presented in Table \ref{table} for this example. Please notice that the third column represents the probability distribution of the raptor code presented in \cite{Raptor}. The rate of the outer code of the A-variable is selected to be 0.25 to encode $k_a=800$ information bits into $n_a=3200$ A-variables. These A-variables are annotated to the 65,536 intermediate bits of the above mentioned raptor code. Simulations show that the average overhead based on the annotation method introduced in this paper is 3.4\%. This amounts to 10\% overhead reduction compared to the average 3.8\% overhead of the original raptor code. We emphasize that the complexity per information bit is exactly the same for both codes. It is worthwhile to mention that by using conventional raptor codes, an overhead of 3.4\% could not be achieved for block lengths less than 80,000 bits \cite{Raptor}, which would involve a much more memory complexity.

\section{Conclusion}
Since raptor codes need a reception overhead to be able to recover the information bits, some of the received bits are indeed never used in the process of decoding. In this work, we presented an extension of the well known raptor codes showing that extra information bits can be embedded through careful annotation of a subset of transmissions. We then detailed the encoding and the decoding process of the proposed codes based on the changes made in the design of the original raptor codes. Finally, we provided a numerical example verifying the improved performance even without optimization a probability distribution for the annotated raptor codes.  Finding the optimal probability distribution for the new encoding/decoding structure will reveal its full potentials.

\begin{table}[t!]
\centering
\renewcommand{\arraystretch}{1.4}
\begin{tabular}{|c|c|c|c|}
\hline
$i$ & $\Phi_{i}$ & $\Psi_{i}$ & $\Omega_{i}=\Phi_{i}+\Psi_{i}$\\
\hline \hline
1 & 0.007969 & 0 & 0.007969\\
\hline
2 & 0.478570 & 0.015 & 0.493570\\
\hline
3 & 0.161220 & 0.005 & 0.166220\\
\hline
4 & 0.072646 & 0 & 0.072646\\
\hline
5 & 0.082558 & 0 & 0.082558\\
\hline
8 & 0.056058 & 0 & 0.056058\\
\hline
9 & 0.037229 & 0 & 0.037229\\
\hline
19 & 0.055590 & 0 & 0.055590\\
\hline
65 & 0.025023 & 0 & 0.025023\\
\hline
66 & 0.003135 & 0 & 0.003135\\
\hline
\end{tabular}
\caption{Example Code with $k=64520$ and $k_a=800$}
\label{table}
\end{table}

\bibliographystyle{IEEEtran}
\bibliography{IEEEabrv,ARC_bib}

\end{document}